\documentclass{JHEP}
\usepackage{amssymb,epsfig}

\newcommand{\be}{\begin{equation}}
\newcommand{\ee}{\end{equation}}

\newcommand{\ssl}{\sl}

\newcommand{\bea}{\begin{eqnarray}}
\newcommand{\eea}{\end{eqnarray}}

\title{Non-Abelian vacua in D=5, N=4  gauged supergravity}

\author{Ali H. Chamseddine\\
Center for Advanced Mathematical Sciences (CAMS)
and Physics Department, Americal University of Beirut,
LEBANON\\ E-mail: \email{chams@aub.edu.lb}}

\author{Mikhail S. Volkov\thanks{Supported
by the DFG grant Wi 777/4-2.}\\
 Institute for Theoretical Physics, Friedrich Schiller
University of Jena, Max-Wien Platz 1, D-07743 Jena, 
GERMANY\\ E-mail: \email{vol@tpi.uni-jena.de}}
 \preprint{FSU TPI 01/01\\CAMS/01-01\\hep-th/0101202}

\abstract{
We study essentially non-Abelian
backgrounds in the five dimensional
N=4 gauged SU(2)$\times$U(1) supergravity.
Static configurations that are invariant under either the SO(4)
spatial rotations or with respect to the
SO(3) rotations and translations
along the fourth spatial coordinate are considered.
By analyzing consistency conditions for the equations for
supersymmetric Killing spinors we derive the Bogomol'nyi
equations and obtain their globally regular solutions.
The SO(4) symmetric configurations contain the purely magnetic
non-Abelian fields together with
the purely electric Abelian field and possess
two unbroken supersymmetries.
The SO(3) configurations have
only the non-Abelian fields and preserve four supersymmetries.}


\begin{document}

\section{Introduction}

The gauged supergravities in five dimensions have been recently the subject
of intensive research in view of the AdS/CFT correspondence (see \cite
{Aharony99} for a review) as well as in connection with the brane world
scenario \cite{Randall99}. It is believed that solutions in such models
provide the dual supergravity description for flat space gauge theories.
This has inspired the widespread interest in such solutions, but only
configurations with Abelian gauge fields have been studied so far. At the
same time, the bulk theories generically contain Yang-Mills fields, which of
course have nothing to do with the non-Abelian fields of the dual gauge
theories but rather give rise to non-trivial warp factors in the
ten-dimensional metric. It would therefore be interesting to obtain
supergravity solutions with non-trivial Yang-Mills fields in the bulk and
implement them in the context of the bulk/boundary correspondence.

Some results in this direction have been obtained in four dimensions. In
\cite{Chamseddine97} the non-Abelian monopole-type supersymmetric vacua were
found in the context of the N=4 half-gauged SU(2)$\times $(U(1))$^{3}$
supergravity of Freedman and Schwarz \cite{Freedman78}, and their
ten-dimensional analogs were obtained in \cite{Chamseddine98}. It has been
argued \cite{Maldacena00} that these solutions provide the dual supergravity
description for the N=1 super-Yang-Mills theory. The non-Abelian Euclidean
supersymmetric backgrounds and their ten-dimensional analogs were obtained
in \cite{Volkov99,Volkov99a}, but the corresponding dual flat space theory
has not been identified so far. Other known solution in D=4 can be related
to reductions of heterotic string theory; see \cite{Gibbons95} and
references therein. The only known non-Abelian vacua in D=5 are the
heterotic solitons of \cite{Gibbons95}, and also the BPS solutions with
non-Abelian matter \cite{Gibbons94}.

In the present paper we study non-Abelian supersymmetric backgrounds in five
dimensions in the context of N=4  SU(2)$\times $U(1) gauged supergravity of
Romans \cite{Romans86}. We consider static configurations that are invariant
either under the SO(4) spatial rotations or with respect to the SO(3)
rotations plus translations along the fourth spatial coordinate. By
analyzing the consistency conditions  for supersymmetric Killing spinors we
derive the Bogomol'nyi equations and obtain their globally regular
solutions. In the SO(4) case the configurations contain the purely magnetic
non-Abelian fields plus the purely electric Abelian field and preserve only
two unbroken supersymmetries out of sixteen. The SO(3) configurations have
only the non-Abelian fields and preserve four supersymmetries.

\section{The D=5, N=4 gauged supergravity}

The five dimensional N=4 gauged SU(2)$\times$U(1) supergravity contains in
the bosonic sector the gravitational field ${\bf g}_{\mu\nu}$, the SU(2)
non-Abelian gauge field $A^a_\mu$ ($a=1,2,3$), the Abelian gauge field $a_\mu
$, a pair of 2-form fields, and the dilaton $\phi$ 
\cite{Romans86}. Since the 2-forms are self-dual, one can set them to zero
on shell, and then one can set the U(1) gauge coupling constant to zero,
such that the model becomes ungauged in the U(1) sector. After a suitable
rescaling of the fields one can set the SU(2) gauge coupling constant to
one, and then the bosonic part of the action becomes
\begin{eqnarray}  \label{1}
S=\int \left(-\frac{R}{4}\right.&+&\frac12\,
\partial_\mu\phi\,\partial^\mu\phi -\frac{1}{4}\,\eta^2
F^{a}_{\mu\nu}F^{a\mu\nu} -\frac{1}{4\eta^4}\,f_{\mu\nu}f^{\mu\nu}  \nonumber
\\
&-&\left.\frac{1}{4\sqrt{{\bf g}}}\,\varepsilon^{\mu\nu\rho\sigma\tau}
F^{a}_{\mu\nu}F^{a}_{\rho\sigma}a_\tau\, +\frac{1}{8\eta^2} \right)\sqrt{%
{\bf g}}\,d^5x\, .
\end{eqnarray}
Here $\eta=\exp(\sqrt{\frac23}\,\phi)$, also $F^a_{\mu\nu}=\partial_\mu
A^a_\nu-\partial_\nu A^a_\mu +\epsilon_{abc}A^b_\mu A^c_\nu$, while the
Abelian field strength is $f_{\mu\nu}=\partial_\mu a_\nu-\partial_\nu a_\mu$.

In the fermionic sector the theory contains four gravitini $\psi _{\mu }^{I}$
and four gaugini $\chi ^{I}$; we shall always omit the index $I=1,\ldots 4$
in what follows. One can set the fermions to zero on shell, however their
SUSY variation in general do not vanish. To write down these variations, let
us introduce $4\times 4$ spacetime gamma matrices $\gamma ^{A}=(\gamma
^{0},\gamma ^{r},\gamma ^{a})$
subject to
\begin{equation}
\gamma ^{A}\gamma ^{B}+\gamma ^{B}\gamma ^{A}=2\eta ^{AB}\,,  \label{2}
\end{equation}
with $\eta_{AB}=(+,-,-,-,-)$, 
and also $4\times 4$ matrices $\Gamma _{j}=(\Gamma _{a},\Gamma _{4},\Gamma
_{5})$ acting on the internal indices of the spinors and spanning the
five-dimensional Euclidean Clifford algebra
\begin{equation}
\Gamma _{i}\Gamma _{j}+\Gamma _{j}\Gamma _{i}=2\delta _{ij}\,.  \label{3}
\end{equation}
Notice that we decompose the five-dimensional tangent space indices as $%
(0,r,a)$, where $r$ takes only one value, `$r$', whereas $a=1,2,3$.
Introducing four sets of Pauli matrices: $\underline{\sigma }^{a}$, $\sigma
_{b}$, $\underline{\tau }^{a}$, and $\tau _{b}$, where matrices from
different sets commute, for example $[\underline{\sigma }^{a},\sigma _{b}]=0$%
, one can choose
\begin{equation}
\gamma ^{0}=\underline{\sigma }^{3}\otimes 1\!{\rm l}_{2},\ \ \ \ \gamma
^{r}=i\underline{\sigma }^{1}\otimes 1\!{\rm l}_{2},\ \ \ \gamma ^{a}=i%
\underline{\sigma }^{2}\otimes \sigma _{a}\,,  \label{4}
\end{equation}
and also
\begin{equation}
\Gamma _{a}=\underline{\tau }^{2}\otimes \tau _{a},\ \ \ \Gamma _{4}=%
\underline{\tau }^{1}\otimes 1\!{\rm l}_{2},\ \ \ \Gamma _{5}=\underline{%
\tau }^{3}\otimes 1\!{\rm l}_{2}\,.  \label{5}
\end{equation}
We shall not write explicitly 
the $\otimes $ symbol and the factors of $1\!{\rm l}_{2}$ in
what follows. One has $\Gamma _{i\ldots j}=\Gamma _{\lbrack i}\ldots
\Gamma _{j]}$, similarly for products of $\gamma ^{A}$. Introducing the
1-form basis $\Theta ^{A}=\Theta _{~\mu }^{A}dx^{\mu }$ such that ${\bf g}%
_{\mu \nu }dx^{\mu }dx^{\nu }=\eta _{AB}\Theta ^{A}\Theta ^{B}$, the
corresponding spin connection is $\omega _{~B}^{A}=\omega
_{~B,\,C}^{A}\Theta ^{C}$. The dual vector basis is defined by $%
E_{A}=E_{A}^{~\mu }\,\partial _{\mu }$ so that the supercovariant derivative
acting on the spinor supersymmetry parameter $\epsilon $ becomes
\begin{equation}
D_{A}\varepsilon =\left( E_{A}^{~\mu }\frac{\partial }{\partial x^{\mu }}+%
\frac{1}{4}\,\omega _{CB,A}\gamma ^{CB}+\frac{1}{2}\,A_{A}^{a}\Gamma
_{a45}\right) \epsilon \,.  \label{6}
\end{equation}
As a result, the linearized SUSY variations of the fermions in the model are
given by \cite{Romans86}
\begin{equation}
\delta \psi _{A}=D_{A}\epsilon +\frac{1}{6\sqrt{2}\eta }\,\gamma _{A}\Gamma
_{45}\,\epsilon -\frac{1}{6\sqrt{2}}\,(\gamma _{A}^{~BC}-4\delta
_{A}^{B}\gamma ^{B})\left( \eta F_{BC}^{a}\Gamma _{a}+\frac{1}{\sqrt{2}\eta
^{2}}\,f_{BC}\right) \epsilon \,,  \label{7}
\end{equation}
\begin{equation}
\delta \chi =\frac{1}{\sqrt{2}}\,\gamma ^{A}(E_{A}^{~\mu }\partial _{\mu
}\phi )\epsilon +\frac{1}{2\sqrt{6}\eta }\,\Gamma _{45}\,\epsilon -\frac{1}{2%
\sqrt{6}}\,\gamma ^{AB}\left( \eta F_{AB}^{a}\Gamma _{a}-\frac{\sqrt{2}}{%
\eta ^{2}}\,f_{AB}\right) \epsilon \,.  \label{8}
\end{equation}

\section{Solutions with SO(4) symmetry}

Our first task is to consider fields which are static and invariant under
the action of the SO(4) spatial symmetry group. The static, SO(4)-invariant
spacetime metric can be represented in the curvature coordinates as
\begin{equation}  \label{9}
ds^2={\rm e}^{2\nu(r)}dt^2- \frac{dr^2}{N(r)}-r^2 d\Omega_3^2\, ,
\end{equation}
where $d\Omega_3^2$ is the round metric of $S^3$. Introducing on $S^3$ the
left-invariant forms $\theta^a$ subject to the Maurer-Cartan equation
\begin{equation}  \label{10}
d\theta^a+\varepsilon_{abc}\,\theta^b\wedge\theta^c=0\, ,
\end{equation}
one has $d\Omega_3^2=\theta^a\theta^a$. The static gauge field $A^a=A^a_\mu
dx^\mu$ that is invariant under the combined action of the SO(4) spatial
rotations and SU(2) gauge transformations is given by
\begin{equation}  \label{11}
A^a=(w(r)+1)\,\theta^a\, ,
\end{equation}
the corresponding field strength being `purely magnetic'
\begin{equation}
F^a=dw\wedge\theta^a+\frac12\,(w^2-1)\,
\varepsilon_{abc}\,\theta^b\wedge\theta^c\,.
\end{equation}
We choose the Abelian field to be `purely electric'
\begin{equation}  \label{12}
f=Q(r)\,dt\wedge dr\, .
\end{equation}
Finally, the dilaton is chosen as $\phi=\phi(r)$. As a result, all fields
are expressed in terms of five functions $\nu$, $N$, $w$, $Q$, $\phi$ of the
radial coordinate $r$.

Varying the action (\ref{1}) gives the second order Lagrangian field
equations. These admit important first integrals. When the five-metric
splits into the direct sum $^{(5)}{\bf g}={\bf g}_{00}\oplus ^{(4)}\!{\bf g}$%
, one can check that not only in the SO(4)-symmetric case but also for
arbitrary static fields, the field equations require that $^{(4)}\nabla \left(
\ln {\bf g}_{00}-2\sqrt{\frac{2}{3}}\phi \right) =0$.  Here $^{(4)}\nabla $ is
the covariant Laplacian with respect to $^{(4)}{\bf g}$. This implies that
the following metric-dilaton relation can be imposed on shell:
\begin{equation}
\nu =\sqrt{\frac{2}{3}}\,(\phi -\phi _{0})\,,  \label{MD}
\end{equation}
where $\phi _{0}$ is an integration constant.

Next, the equations for the Abelian field $f$
\begin{equation}  \label{13}
\nabla_\nu(\xi^{-4}f^{\nu\mu})=\frac{1}{4\sqrt{{\bf g}}}\,
\varepsilon^{\mu\nu\rho\sigma\tau}F^a_{\nu\rho}F^a_{\sigma\tau}
\end{equation}
have the total derivative structure. In the SO(4)-symmetric case 
they can be integrated to
give
\begin{equation}  \label{14}
Q=\frac{{\rm e}^{5\nu}}{\sqrt{N}r^3}\,(2w^3-6w+H),
\end{equation}
with $H$ being integration constant. The remaining independent Lagrangian
equations read
\begin{eqnarray}
\frac{r^3}{2}N^{\prime}+r^2(N-1) +r^2N{\rm e}^{2\nu}w^{\prime 2} &+&{\rm e}%
^{2\nu}(w^2-1)^2 +\frac{r^4}{2}N\nu^{\prime 2}  \nonumber \\
&-&\frac{r^4}{12}{\rm e}^{-2\nu} +\frac13 r^4N{\rm e}^{-6\nu}Q^2=0\, ,
\nonumber \\
\frac{r^3}{2}N^{\prime}+ 2r^2N{\rm e}^{2\nu}w^{\prime 2} &+&r^4N\nu^{\prime
2}-r^3N\nu^{\prime}=0\, , \\
r^2Nw^{\prime\prime}+(3r^2N\nu^{\prime}+rN+\frac{r^2}{2}N^{\prime})w^{%
\prime}&-&2r{\rm e}^{-3\nu}\sqrt{N}(w^2-1)Q =2(w^2-1)w\, .  \nonumber
\label{15}
\end{eqnarray}
There is also an equation containing $\nu^{\prime\prime}$, but it can be
related to the equations above by virtue of the Bianchi identities.

\subsection{Supersymmetry constraints}

Our aim now is to study constraints imposed by supersymmetry. These can be
expressed as a system of linear differential equations for the spinor
supersymmetry parameters, $\delta\psi_A=\delta\chi=0$. These equations are
generically inconsistent, however one can find the consistency conditions,
which can be represented as a set of non-linear first order differential
equations for the background variables; see Eqs.(\ref{32}). These equations,
usually called Bogomol'nyi equations, are further first integrals for the
Lagrangian field equations.

Let us introduce the 1-form basis
\begin{equation}  \label{16}
\Theta^0={\rm e}^{\nu}dt,\ \ \ \Theta^r=\frac{dr}{\sqrt{N}},\ \ \
\Theta^a=r\,\theta^a\, ,
\end{equation}
such that the spacetime metric is
\begin{equation}  \label{17}
ds^2=(\Theta^0)^2-(\Theta^r)^2-\delta_{ab}\,\Theta^a\Theta^b\, .
\end{equation}
The dual vielbein vectors $E_A$ are
\begin{equation}  \label{18}
E_0={\rm e}^{-\nu}\frac{\partial}{\partial t}, \ \ E_r=\sqrt{N}\frac{\partial%
}{\partial r}, \ \ E_a=\frac{1}{r}\, e_a\, .
\end{equation}
Here $\theta^a$ are the left-invariant Maurer-Cartan forms on $S^3$ subject
to Eq.(\ref{10}), and $e_b$ are the dual left-invariant vectors, $%
\langle\theta^a,e_b\rangle=\delta^a_b$. It is worth noting that $e_a$,
together with the right-invariant vectors, $\tilde{e}_a$, give rise to the
angular momentum operators $L_a=\frac{i}{2}e_a$ and $\tilde{L}_a=\frac{i}{2}%
\tilde{e}_a$ with the commutation relations
\begin{equation}  \label{19}
[L_a,L_b]=i\varepsilon_{abc}\,L_c\, ,~~~~ [\tilde{L}_a,\tilde{L}%
_b]=i\varepsilon_{abc}\,\tilde{L}_c\, ,~~~~ [L_a,\tilde{L}_b]=0\, .
\end{equation}
One also has $L_aL_a=\tilde{L}_a\tilde{L}_a$. The spin-connection is given
by $${\omega}_{AB,C}=\frac12\,(C_{B,AC}+C_{C,AB}-C_{A,BC})$$ where $C_{A,BC}={%
\eta}_{AD}C^D_{\ BC}$ are determined by the commutation relations for the
basis vectors of the vielbein, $[{E}_A,{E}_B]=C^C_{\ AB}\,{E}_C$. One finds
the following non-zero components:
\begin{equation}  \label{20}
\omega_{0r,0}=\sqrt{N}\nu^{\prime},\ \ \ \omega_{ra,b}=\frac{\sqrt{N}}{r}%
\,\delta_{ab},\ \ \ \omega_{ab,c}=\frac{1}{r}\,\varepsilon_{abc}\, .\ \ \
\end{equation}

Inserting the above expressions into (\ref{7}),(\ref{8}) and assuming that
all spinors are time-independent, we compute the spinor SUSY variations $%
\delta \chi $ and $\delta \psi _{A}$. First, we obtain
\begin{equation}
\delta \chi =\sqrt{3}\,\underline{\sigma }^{3}\,\delta \psi _{0}-\frac{i%
\sqrt{3}}{2r}\left( \nu -\sqrt{\frac{2}{3}}\phi \right) ^{\prime }\underline{%
\sigma }^{1}\,\epsilon \,,  \label{21}
\end{equation}
which implies, in view of the metric-dilaton relation (\ref{MD}), that $%
\delta \chi $ is not an independent variation. We therefore focus on the
gravitino variations $\delta \psi _{A}$:
\begin{eqnarray}
\delta \psi _{0} &=&\left( -\frac{1}{2}{\cal A}_{1}\,\underline{\sigma }^{2}-%
\frac{i}{2}{\cal A}\,\underline{\sigma }^{3}\underline{\tau }^{2}+\frac{i}{6}%
\,({\cal C}+{\cal B}\,\underline{\sigma }^{3}\,)\,\underline{\tau }%
^{2}\,(\sigma ^{a}\tau ^{a})+i{\cal F}\,\underline{\sigma }^{1}\right)
\epsilon \,,  \nonumber  \label{22} \\
\delta \psi _{r} &=&\left( \sqrt{N}\frac{\partial }{\partial r}-\frac{1}{2}%
{\cal A}\,\underline{\sigma }^{1}\underline{\tau }^{2}+\frac{1}{6}\,(2i{\cal %
C}\,\underline{\sigma }^{2}\,+{\cal B}\,\underline{\sigma }^{1}\,)\,%
\underline{\tau }^{2}\,(\sigma ^{a}\tau ^{a})-{\cal F}\,\underline{\sigma }%
^{3}\right) \epsilon \,,  \nonumber \\
\delta \psi _{a} &=&\left( -\frac{2i}{r}\,L_{a}-\frac{i}{2}\,{\cal B}_{1}\,%
\underline{\sigma }^{3}\sigma _{a}-\frac{i}{2r}\,(\sigma _{a}+\tau _{a})-%
\frac{i}{2}\,{\cal C}_{1}\tau _{a}\right.   \nonumber \\
&-&\frac{1}{2}{\cal A}\,\underline{\sigma }^{2}\underline{\tau }^{2}\sigma
_{a}+\frac{1}{6}\left. \,({\cal B}\,\underline{\sigma }^{2}\Sigma _{a}\,-%
{\cal C}\,\underline{\sigma }^{1}\Lambda _{a})\,\underline{\tau }^{2}-\frac{i%
}{2}\,{\cal F}\,\sigma _{a}\right) \epsilon \,.
\end{eqnarray}
Here $\Sigma _{a}=\tau _{a}+2i\,\varepsilon _{abc}\sigma _{b}\tau _{c}$, $%
\Lambda _{a}=2i\,\tau _{a}-\varepsilon _{abc}\sigma _{b}\tau _{c}$, and also
the following abbreviations have been introduced:
\begin{eqnarray}
{\cal A} &=&\frac{{\rm e}^{-\nu }}{3\sqrt{2}},\ \ {\cal B}=\sqrt{2}{\rm e}%
^{\nu }\frac{w^{2}-1}{r^{2}},\ \ {\cal C}=\frac{\sqrt{2N}}{r}{\rm e}^{\nu
}w^{\prime },  \nonumber  \label{22a} \\
{\cal A}_{1} &=&\sqrt{N}\nu ^{\prime },\ \ {\cal B}_{1}=\frac{\sqrt{N}}{r},\
\ {\cal C}_{1}=\frac{w}{r}\,,\ \ \ {\cal F}=\frac{{\rm e}^{2\nu }}{3r^{3}}%
\,(2w^{3}-6w+H).
\end{eqnarray}
The supersymmetry constraints are obtained by setting $\delta \psi _{A}=0$,
which gives the system of equations for the spinor $\epsilon $. This spinor
has 16 complex components subject to the symplectic Majorana condition, such
that there are altogether 16 real independent components. Let us introduce
two component spinors of four different types, $\underline{\psi }$, $\psi $,
$\underline{\xi }$, $\xi $, that live in four spinor spaces where the
operators $\underline{\sigma }^{a}$, $\sigma _{b}$, $\underline{\tau }^{a}$,
and $\tau _{b}$, respectively act. One can expand $\epsilon $ as
\begin{equation}
\epsilon =\sum_{\alpha ,\beta \,\gamma ,\delta =\pm 1}C_{\alpha \beta \gamma
\delta }\,\,\underline{\psi }^{\alpha }\otimes \psi ^{\beta }\otimes
\underline{\xi }^{\gamma }\otimes \xi ^{\delta }\,,
\end{equation}
where $C_{\alpha \beta \gamma \delta }$ are 16 functions of spacetime
coordinates, and $\underline{\sigma }^{3}\underline{\psi }^{\alpha }=\alpha
\underline{\psi }^{\alpha }$, also ${\sigma }^{3}{\psi }^{\beta }=\beta \psi
^{\beta }$ and ${\tau }^{3}{\xi }^{\delta }=\delta \xi ^{\delta }$, while $%
\underline{\xi }^{\gamma }$ are chosen to be eigenvectors of $\underline{%
\tau }^{2}$, $\underline{\tau }^{2}\underline{\xi }^{\gamma }=\gamma
\underline{\xi }^{\gamma }$. The supersymmetry constraints $\delta \psi
_{A}=0$ is a system of 5$\times $16=80 equations for 16 components of $%
\epsilon $. Coefficients of this system, defined in (\ref{22a}), are
determined by the underlying bosonic configuration. Although generically
only the trivial solution is possible, one can find consistency conditions
for the coefficients under which non-trivial solutions exist as well. The
first step in doing this is to reduce somehow the size of the system. Since
the underlying configuration is SO(4)-invariant, it is natural to consider
the sector where $\epsilon $ is the eigenstate of the SO(4) angular momentum
with zero eigenvalue(s).

Since SO(4) is locally isomorphic to the product of two copies of SO(3), the
SO(4) angular momentum is essentially the sum of two SO(3) angular momenta.
The two commuting SO(3) orbital angular momentum operators are given by Eq.(%
\ref{19}), but since the fermions also carry spin and isospin, we need the
operator of the total angular momentum:
\begin{equation}  \label{24}
J_a=L_a+\frac12\,(\sigma_a+\tau_a)\, .
\end{equation}
Since the commuting operators are $J^2$, $J_3$, $\tilde{L}^2$, $\tilde{L}_3$%
, there is a spinor $\epsilon$ annihilated by all these operators, such that
$J_a\epsilon=\tilde{L}_a\epsilon$=0, and in view of the relation $L^2=\tilde{%
L}^2$ one has also $L_a\epsilon=0$, which implies that
\begin{equation}  \label{25}
L_a\epsilon=0,\ \ \ \ (\sigma_a+\tau_a)\epsilon=0\, .
\end{equation}
The solution of these equations is
\begin{equation}  \label{25a}
\epsilon=(\psi^{+1}\xi^{-1}-\psi^{-1}\xi^{+1})\sum_{\alpha,\gamma=\pm1 }
C_{\alpha\gamma}(r)\,\, \underline{\psi}^\alpha\underline{\xi}^\gamma\, ,
\end{equation}
and so we are now left with only four independent unknown functions $%
C_{\alpha\gamma}(r)$. From three matrices $\underline{\tau}^a$ only $%
\underline{\tau}^2$ enters the SUSY variations (\ref{22}) and this leaves
subspaces generated by $\underline{\xi}^{+1}$ and $\underline{\xi}^{-1}$
invariant. As a result, inserting (\ref{25a}) into (\ref{22}) and denoting $%
\Psi_\gamma=\sum_{\alpha=\pm1}C_{\alpha\gamma}(r) \underline{\psi}^\alpha$,
the equations for $\Psi_{+1}$ decouple from those for $\Psi_{-1}$. The
conditions $\delta\psi_0=0$ and $\delta\psi_a=0$ reduce then to
\begin{equation}  \label{26}
\left({\cal A}_1\,\underline{\sigma}^3 -i\gamma\, {\cal A}\underline{\sigma}%
^2 +\gamma\,{\cal C}\underline{\sigma}^1 -i\gamma\,{\cal B}\underline{\sigma}%
^2 -2{\cal F} \right)\Psi_\gamma=0\, ,
\end{equation}
\begin{equation}  \label{27}
\left({\cal B}_1\,\underline{\sigma}^3 -i\gamma\,{\cal A} \underline{\sigma}%
^2 -{\cal C}_1 +i\gamma\,{\cal B}\underline{\sigma}^2 +{\cal F}
\right)\Psi_\gamma=0\, ,
\end{equation}
while $\delta\psi_r=0$ gives
\begin{equation}  \label{27a}
\left(\sqrt{N}\frac{d}{dr} -\frac{\gamma}{2}\,({\cal A}+{\cal B})\,
\underline{\sigma}^1 -i\gamma\,{\cal C}\,\underline{\sigma}^2 -{\cal F}%
\underline{\sigma}^3 \right)\Psi_\gamma=0\, .
\end{equation}
Let us first consider Eqs.(\ref{26}),(\ref{27}). For a given $\gamma=\pm 1$
these are four homogeneous algebraic equations for the two unknown
quantities $C_{+1\gamma}$ and $C_{-1\gamma}$. A nontrivial solution exists
if the 4$\times$2 matrix of the system has rank 1, which gives three
conditions on the coefficients of the matrix:
\begin{eqnarray}
{\cal A}_1^2+{\cal C}^2=({\cal A}+{\cal B})^2+4{\cal F}^2\, ,  \nonumber \\
{\cal B}_1^2=({\cal F}-{\cal C}_1)^2+({\cal A}-{\cal B})^2\, ,  \nonumber \\
({\cal A}-{\cal B})({\cal C}-{\cal A}-{\cal B})= ({\cal A}_1-2{\cal F})(%
{\cal F}-{\cal B}_1-{\cal C}_1) .  \label{28}
\end{eqnarray}
Notice that these relations do not contain $\gamma$. Under these conditions
the algebraic equations (\ref{26}),(\ref{27}) become consistent and admit
two solutions
\begin{equation}  \label{29}
C_{-1\gamma}(r) =\gamma\,\frac{2{\cal F}-{\cal A}_1} {{\cal C}-{\cal A}-%
{\cal B}}\,C_{+1\gamma}(r) \equiv \gamma {\cal Q} \,C_{+1\gamma}(r)
\end{equation}
corresponding to two different values of $\gamma$.

Now, inserting these solutions into the differential constraints (\ref{27a})
gives two linear first order differential equations for one function $%
C_{+1+1}(r)$, and the same pair of equation arises for $C_{+1-1}(r)$. Since
two differential equations for the same function must be compatible, this
gives a further constraint on the coefficients:
\begin{equation}  \label{30}
\sqrt{N}{\cal Q}^{\prime}+2{\cal AF}= \frac{{\cal A}+{\cal B}}{2}-{\cal C}- (%
\frac{{\cal A}+{\cal B}}{2}+{\cal C}){\cal Q}^2\, .
\end{equation}
It turns out however that this new constraint is fulfilled by virtue of Eqs.(%
\ref{28}) (checking of which is not completely trivial). The differential
equations can now be solved to give
\begin{equation}  \label{31}
C_{+1\gamma}(r)=C_\gamma\exp\left(\int^r\frac{dr}{\sqrt{N}}\, \{{\cal F}+(%
\frac{{\cal A}+{\cal B}}{2}+{\cal C})\,{\cal Q}\}\right)\, ,
\end{equation}
where $C_\gamma$ are two integration constants. This finally gives two
non-trivial solutions for supersymmetry Killing spinors. The consistency
conditions for the existence of these solutions are given by Eqs.(\ref{28}).

\subsection{Bogomol'nyi equations}

Summarizing the results of the preceding subsection, Eqs.(\ref{28}) contain
the complete set of consistency conditions under which non-trivial solutions
for supersymmetry Killing spinors exist. These conditions can be represented
as a system of first order Bogomol'nyi equations for the background
variables:
\begin{eqnarray}  \label{32}
N&=&\left(\frac{1}{3}\,\xi^2V-w\right)^2+2\xi^2(w^2-1)^2 -\frac23\,(w^2-1)+%
\frac{1}{18\xi^2}\, ,  \nonumber \\
r\,\frac{dw}{dr}&=& \frac{1}{6\xi^2N}\left\{2V(1-w^2)\,\xi^4+(H-4w^3)\,%
\xi^2-w\right\}\, ,  \nonumber \\
r\,\frac{d\xi}{dr}&=&-\frac{\xi}{3N} \left\{V^2\xi^4+(12\,(w^2-1)^2-4Vw)\,%
\xi^2+w^2+2\right\}\, ,
\end{eqnarray}
with $\xi={\rm e}^\nu/r$ and $V=2w^3-6w+H$. One can directly check that
these equations are compatible with the Lagrangian equations of motion (\ref
{15}). Any solution to the Bogomol'nyi equations preserves {\it two}
supersymmetries.

One can obtain some simple solutions. For example, setting $H=0$, we find
that $w=0$ is a solution. The corresponding geometry
\begin{equation}
ds^{2}=r_{0}^{2}\,{\rm e}^{\frac{1}{12\xi ^{2}}}\left( \xi \,dt^{2}-\frac{1}{%
8\xi ^{5}}\,d\xi ^{2}-\frac{1}{\xi }\,d\Omega _{3}^{2}\right)   \label{33}
\end{equation}
is singular both at the origin of the spherical coordinate system and at
infinity (here $r_{0}$ is the integration constant).

The geometry of the solutions can be regular at the origin, $r= 0$, if only $%
H=4$. Introducing the new variable $Y=\frac{1}{\xi^2}+2w^2+4w-2-\frac{4}{w}$%
, the Bogomol'nyi equations (\ref{32}) reduce then to
\begin{equation}  \label{34}
w^2Y\,\frac{dY}{dw}=4\,(w-1)^2Y +16\,(w-1)(2w+1)(w+2).
\end{equation}
Some solutions to this Abel's equation are known in a closed analytical
form: $Y=4(2w+1)(w-1)/w$ and $Y=-2(2w+1)(w+2)/w$, which however give rise to
$\xi^2<0$. The numerical analysis on the other hand reveals a smooth
solution with the following asymptotics (see Fig.1):
\begin{eqnarray}
Y&=&8+4\cdot 7\,w+4\cdot 23\,w^2+8\cdot 89\,w^3+12\cdot 157\,w^4+\ldots ~~%
{\rm as}~w\to 0\, ;~~~  \nonumber \\
Y&=&12\,x+4\,x^2 +2\,x^3+\frac{14}{15}\,x^4 +\frac{3}{10}\,x^5-\frac{23}{210}%
\,x^6+\ldots~~{\rm as}~w\to 1\, ,  \label{as}
\end{eqnarray}
here $x=1-w$. The appearance of the prime numbers in these expansions
suggests that the analytical solution with such asymptotics, if exists, 
should be sought for in a parametric form rather then as $Y(w)$.
Passing to the $w(r)$, $N(r)$, $\nu(r)$ parameterization of this solution one
finds that the geometry is globally regular; see 
Fig.2. At the origin one has $w=1+O(r^2)$, $%
N=1+O(r^2)$, $\nu=O(r^2)$, such that the curvature is bounded and the gauge
field vanishes as $r\to 0$. At infinity, $r\to\infty$, the leading behavior
of the field amplitudes is $N\sim 1/w\sim r{\rm e}^{-\nu}\sim \ln r$, such
that the geometry is not asymptotically flat (and not asymptotically AdS).

\begin{figure}[h]
\begin{minipage}[b]{0.45\linewidth}
  \centering\epsfig{figure=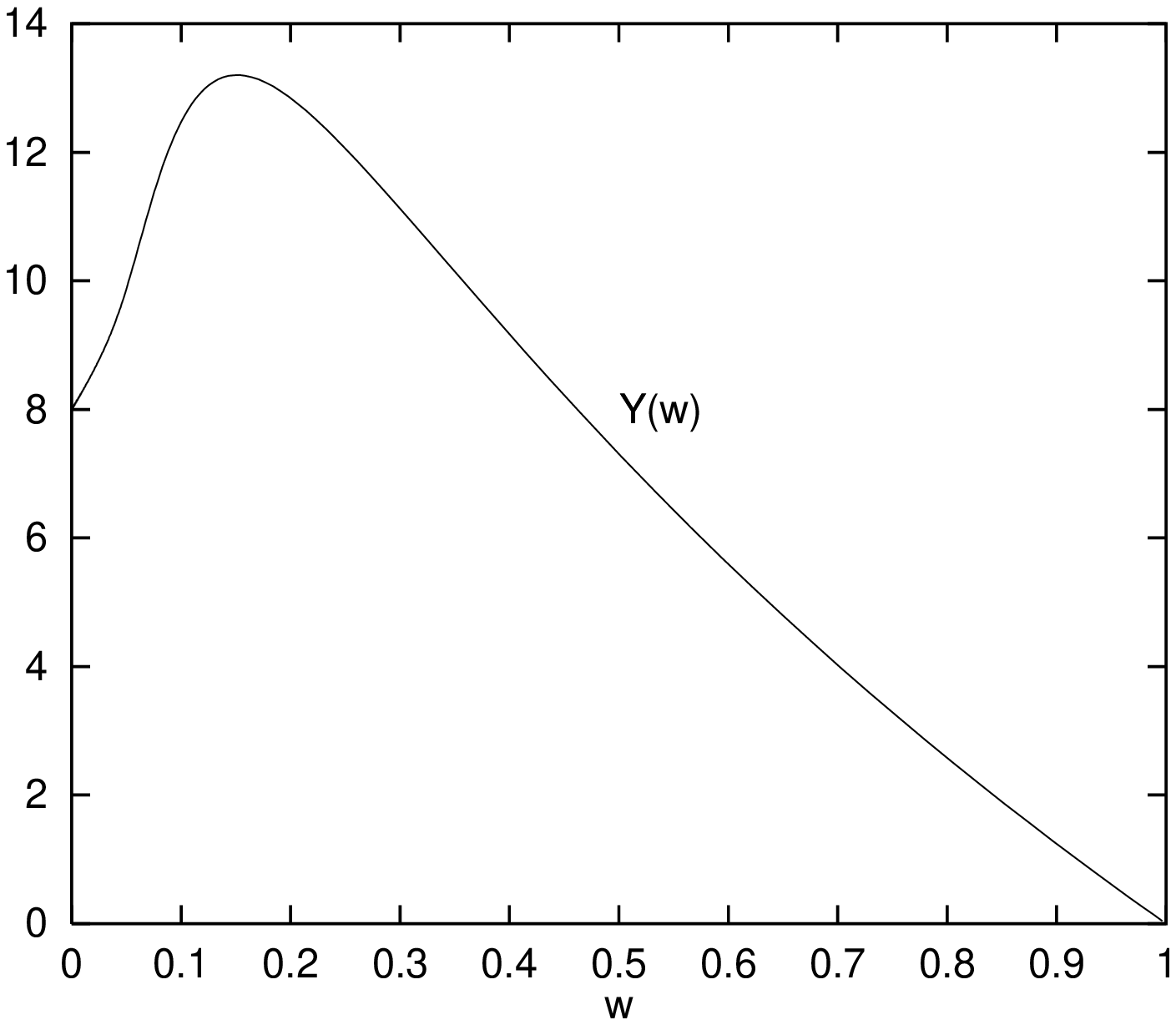,width=1.2\linewidth}
  \caption{{\fontsize{10}{12}\selectfont 
Solution to the Bogomol'nyi equation (\ref{34}) with the 
boundary conditions specified by  Eq.(\ref{as}).}}
  \label{fig1}
   \end{minipage}\hspace{4 mm}
\begin{minipage}[b]{0.45\linewidth}
  \centering\epsfig{figure=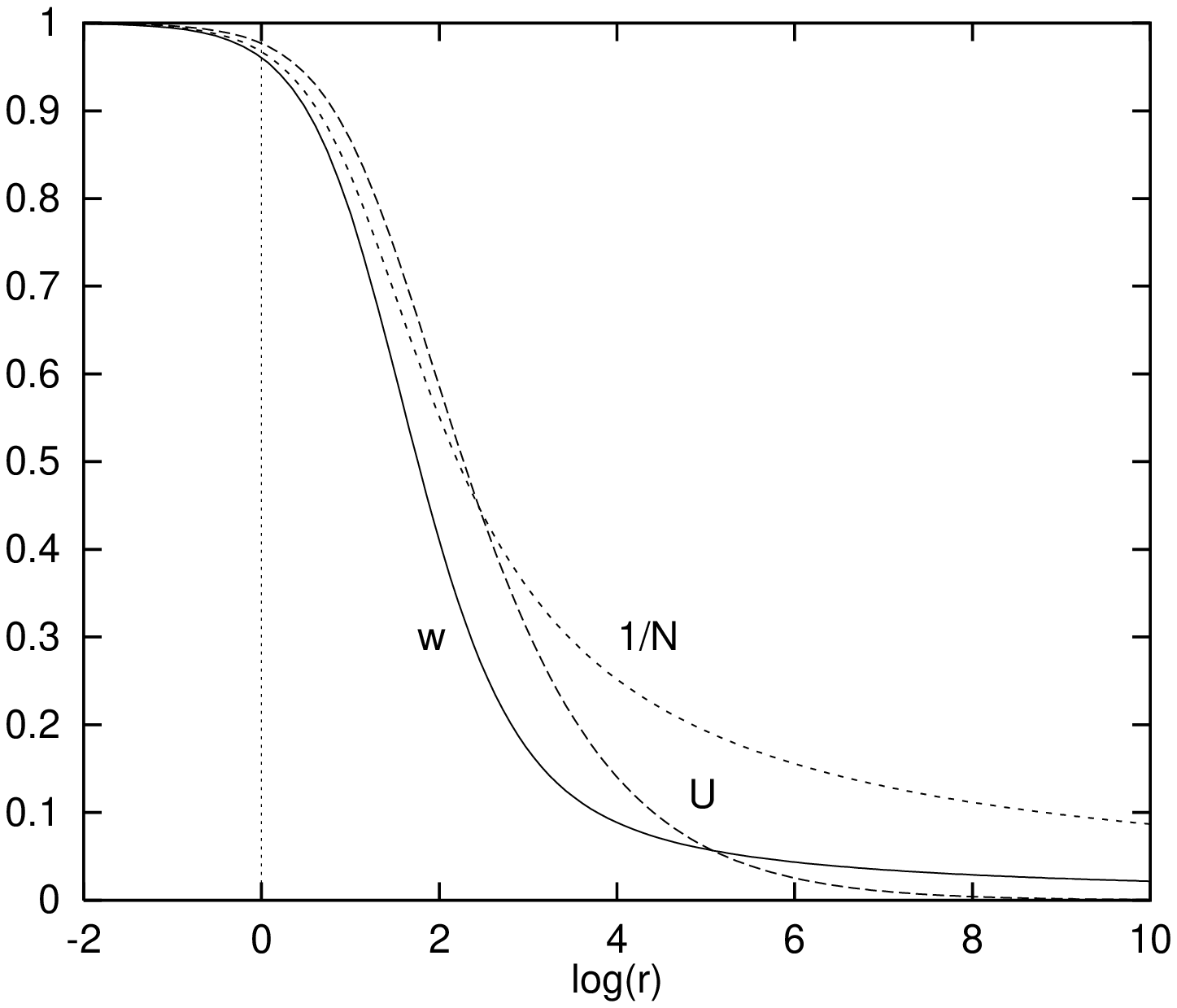,width=1.2\linewidth}
  \caption{{\fontsize{10}{12}\selectfont
The same solution as in Fig.1 parameterized by $w(r)$,  
$N(r)$, $\nu(r)$ such that Eqs.(\ref{32}) are fulfilled.  
 Here
U$\equiv\exp(-\nu)$. 
}}
  \label{fig2}
   \end{minipage}
\end{figure}

\section{Solutions with SO(3) symmetry}

Our next task is to consider static fields that are invariant under the
SO(3) spatial rotations and in addition under translations along the fourth
spatial direction. We parameterize the metric as
\begin{equation}  \label{40}
ds^2={\rm e}^{2\nu}\,dt^2-{\rm e}^{2\tau}\left(\frac{d\tau^2}{N}+
d\Omega_2^2\right) -{\rm e}^{2\mu}\,(dx^4)^2\, ,
\end{equation}
where $d\Omega_2^2=d\vartheta^2+\sin^2\vartheta d\varphi^2$, and choose the
gauge field according to
\begin{equation}  \label{41}
A^a\mbox{\bf T}_a=w\,(-{\bf T}_2\,d\vartheta+ {\bf T}_1\,\sin\vartheta\,d%
\varphi)+ {\bf T}_3\,\cos\vartheta\,d\varphi\, .
\end{equation}
Here $\nu$, $N$, $\mu$, $w$, and the dilaton $\phi$ depend only on $\tau$,
and $[\mbox{\bf T}_a,\mbox{\bf T}_b]=i\varepsilon_{abc}\mbox{\bf T}_c$ are
the gauge group generators. This gauge field is `purely magnetic', and
moreover its field strength is such that $\varepsilon^{\mu\nu\rho\sigma\tau}
F^{a}_{\mu\nu}F^{a}_{\rho\sigma}=0$. As a result, the Abelian vector field
decouples, and we can set it to zero.

Our strategy is very much similar to the one described above for the
SO(4)-symmetric fields. For this reason we shall mention only the essential
points. First, it turns out that the Lagrangian equations of motion allow us
to impose on-shell the `metric-dilaton relations'
\begin{equation}  \label{MD1}
\nu=\mu-\mu_0=\sqrt{\frac23}\,(\phi-\phi_0)\,
\end{equation}
similar to the one in Eq.(\ref{MD}), $\mu_0$ and $\phi_0$ being integration
constants. The remaining independent equations read
\bea                        \label{42}
\frac32\,N'&-&9N+1-6N\,\frac{\xi'}{\xi}
+10\,N\xi^2w'^2+2N\,\frac{\xi'^2}{\xi^2}
+\frac{1}{2\xi^2}=0\, , \nonumber \\
\frac32\,N'&-&1+2\xi^2(w^2-1)^2+6N\xi^2w'^2
+N\,\frac{\xi'^2}{\xi^2}=0\, , \nonumber \\
Nw''&+&\left(\frac{N'}{2}+3N+4N\frac{\xi'}{\xi}\right)w'=w^3-w\, ,
\eea
with $^\prime=\frac{d}{d\tau}$. The next step is to study the supersymmetry
constraints $\delta\chi=\delta\psi_A=0$ to derive the Bogomol'nyi equations.
Let us split the tangent space indices as $A=(0,\tau,2,3,4)$. It turns out
that the metric-dilaton relation (\ref{MD1}) implies that $\delta\psi_4$ and
$\delta\chi$ fermionic SUSY variations are not independent but can be
expressed in terms of $\delta\psi_0$ via a relation similar to the one in
Eq.(\ref{21}). As a result, the independent supersymmetry constrains are $%
\delta\psi_0=\delta\psi_2=\delta\psi_3=0$, and also $\delta\psi_\tau=0$,
which gives a system of 64 equations.

In order to truncate the system, we require that $J_a\epsilon=0$. Here $%
J_a=L_a+\frac12\,(\sigma_a+\tau_a)$ is the total angular momentum with $L_a$
being the usual SO(3) angular momentum acting on the $\vartheta$, $\varphi$
variables. Since now $L^2$ does not commute with $J_a$, we cannot require
that $\epsilon$ is annihilated separately by the operators $L_a$ and $%
\frac12\,(\sigma_a+\tau_a)$, as was possible in the SO(4) case, but only by
their sum. As a result, $\epsilon$ is constructed in terms of tensor
products of eigenfunctions of $L_3$ and those of $\frac12\,(\sigma_3+\tau_3)$
with eigenvalues $0,\pm 1$. For more details we refer to \cite{Volkov99a}
where a similar problem in four spacetime dimensions was considered.

The resulting ansatz for $\epsilon$ fixes the angular dependence of spinors
and reduces the $\delta\psi_0=\delta\psi_2=\delta\psi_3=0$ constraints to a
system of algebraic equations, whose consistency conditions are obtained
similarly as was done above. These consistency conditions can be represented
as a system of Bogomol'nyi equations,
\begin{eqnarray}  \label{43}
N&=&\frac{S^2}{18\xi^2P} \, ,  \nonumber \\
w^{\prime}&=&\frac{3w}{S}\,(1+2\xi^2(w^2-1))\, ,  \nonumber \\
\xi^{\prime}&=&-\frac{6\xi^3}{S}\,(1+w^2+2\xi^2(w^2-1)^2)\, ,
\end{eqnarray}
with $S=4(w^2-1)^2\xi^4+4(w^2+1)\xi^2+1$, $P=8(w^2-1)^2\xi^4+6(w^2+1)\xi^2+1$%
, and $\xi=\exp(\nu-\tau)$. One can check that these Bogomol'nyi equations
are compatible with the Lagrangian equation (\ref{42}). The remaining $%
\delta\psi_\tau=0$ constraint equations turn out to be compatible with each
other by virtue of Eqs.(\ref{42}), and they completely specify the $\tau$%
-dependence of the spinors. This finally gives {\it four} independent
supersymmetry Killing spinors.

Introducing $Y=1/(2\xi ^{2})$ and $x=w^{2}$, the problem of solving the
Bogomol'nyi equations (\ref{43}) reduces to one equation
\begin{equation}
x(Y+x-1)\,\frac{dY}{dx}+(x+1)Y+(x-1)^{2}=0\,.  \label{44}
\end{equation}
For reasons that will be explained shortly, this equation exactly coincides
with the one previously obtained \cite{Chamseddine97} in the context of the
four-dimensional gauged supergravity of Freedman and Schwarz \cite
{Freedman78}. With the substitution \cite{Chamseddine97}
\begin{equation}
x=\rho ^{2}\,e^{\xi (\rho )},\ \ \ \ \ \ \ Y=-\rho \frac{d\xi (\rho )}{d\rho
}-\rho ^{2}\,e^{\xi (\rho )}-1,  \label{45}
\end{equation}
Eq.(\ref{44}) reduces to the Liouville equation
\begin{equation}
\frac{d^{2}\xi }{d\rho ^{2}}=2\,e^{\xi }\,,  \label{46}
\end{equation}
which is completely integrable. This leads to the following most general
solution of the Bogomol'nyi equations that is regular at the origin of the
spherical coordinate system:
\begin{equation}
ds^{2}=r_{0}^{2}\,{\rm e}^{2\nu }\left\{ dt^{2}-d\rho ^{2}-Y\,d\Omega
_{2}^{2}-(dx^{4})^{2}\right\} \,,  \label{48}
\end{equation}
where $r_{0}$ is the integration constant and
\begin{equation}
Y=2\rho \coth \rho -\frac{\rho ^{2}}{\sinh ^{2}\rho }-1,\ \ \ w=\frac{\rho }{%
\sinh \rho }\,,\ \ \ {\rm e}^{6\nu }=\frac{\sinh ^{2}\rho }{Y}\,,  \label{47}
\end{equation}
while the gauge filed and the dilaton are given by (\ref{41}) and (\ref{MD1}%
). 
Since $Y(\rho )=\rho ^{2}+O(\rho ^{4})$ for small $\rho $, the geometry is
regular as $\rho \rightarrow 0$. The geometry is also globally regular,
although, since $Y=2\rho +O(1)$ as $\rho \rightarrow \infty $, the metric
does not become flat for large $\rho $.

This five-dimensional solution is closely related to the solution of 
the  gauged $D=4$ supergravity of Freedman and Schwarz because 
the latter can be obtained via dimensional reduction
plus truncation of the five-dimensional supergravity under 
consideration. In other words, the five dimensional solution can also 
 be obtained by uplifting the four dimensional solution. 
The relation between the vielbeins in four and five
dimensions is
$\Theta^A={\rm e}^{-\frac13\tilde{\phi}}e^A$, where $A=0,1,2,3$ and 
$e^A$ is the D=4 tetrad, while 
$\Theta^4={\rm e}^{\frac23\tilde{\phi}}dx^4$. The four dimensional dilaton, 
$\tilde{\phi}$, is related to the five dimensional one via 
$\phi =\sqrt{\frac{2}{3}}\,\tilde{\phi}$.  The four-dimensional 
Yang-Mills field is obtained from the five-dimensional one
by setting the fourth spacetime component to zero.

\section{Concluding remarks}

One can lift the above solutions to ten dimensions using the results 
of \cite{Lu99}. The bulk/boundary interpretation of the SO(3) solutions 
will then probably be similar to that 
 for their D=4 counterparts \cite{Maldacena00}
-- they will 
provide the dual supergravity description for the NS 5-branes wrapped 
around $S^2$. It is less clear what the interpretation for the SO(4)
solutions might be. Notice that these solutions do not have
a simple asymptotic behavior -- they do not approach the maximal
(super)-symmetry backgrounds at infinity. This is due to the fact that we
actually consider the half-gauged model, in which case the dilaton potential
has no stationary points thus driving the dilaton asymptotically to
infinity. Turning on the U(1) gauge coupling constant $g_1$ the potential
becomes \cite{Romans86}
\begin{equation}
U(\phi)=-\frac{1}{8}\exp(-2\sqrt{\frac23}\phi) -\frac{g_1}{2\sqrt{2}}\exp(%
\sqrt{\frac23}\phi),
\end{equation}
and this does have a stationary point. This suggests that there could be
asymptotically AdS solutions. In fact some of such solutions have
 recently been obtained \cite{Nieder00}. 
The problem however is that unless $g_1=0$,
the simple metric-dilaton relations as those in (\ref{MD}), (\ref{MD1}) do
not hold and there is no linear dependence between different components of
the fermionic SUSY variations similar to the one in (\ref{21}). This gives
too many independent supersymmetry constraints, which will probably kill all
supersymmetric solutions apart from the simplest ones (all solutions of
\cite{Nieder00} are simple in the sense that they belong to the 
embedded Abelian type). 
 However, a further analysis is required in order to make any definite
statements.


\end{document}